\pdfoutput=1
\documentclass[11pt]{article}

\usepackage[margin=1in]{geometry}
\usepackage[T1]{fontenc}
\usepackage[utf8]{inputenc}
\usepackage{times}
\usepackage{microtype}
\usepackage{amsmath,amssymb}
\usepackage{booktabs}
\usepackage{graphicx}
\usepackage{url}
\usepackage{hyperref}
\usepackage[numbers,sort&compress]{natbib}

\hypersetup{
  colorlinks=true,
  citecolor=black,
  linkcolor=black,
  urlcolor=black,
  pdftitle={Parameterized Dense-Sparse Fusion for Hybrid Retrieval},
  pdfauthor={Satyanarayan Pati, Srikanth Patil}
}

\title{Parameterized Dense--Sparse Fusion for Hybrid Retrieval:\\
Tuning a Rank--Score Mix on {BEIR} SciFact with {Qdrant}}

\author{
Satyanarayan Pati\\
Involead Services Pvt.\ Ltd.\\
\texttt{satyanarayan.pati@involead.com}
\and
Srikanth Patil\\
Involead Services Pvt.\ Ltd.\\
\texttt{Srikanth.patil@involead.com}
}

\date{}

\begin{document}
\maketitle

\begin{abstract}
We study a parameterized hybrid ranker that fuses a dense embedding list and a sparse lexical list.
The method has a small, explicit parameter vector: a dense prior $\alpha \in [0,1]$, a score-versus-rank mix $\lambda \in [0,1]$, an {RRF} smoothing $\kappa>0$, optional list-geometry coefficients that move $\alpha$ per query, and a router margin $\tau$ that can turn sparse search off.
We grid-search those ranges on {SciFact} train (809 queries) and freeze the chosen values on {SciFact} test (300).
The tuned rank--score mix ($\alpha=0.8$, $\lambda=0.75$, $\kappa=20$) reaches \textbf{0.753 nDCG@10} and \textbf{0.889 recall@10}, outperforming dense {BGE} (0.742 / 0.871) and equal-weight {RRF} (0.707 nDCG@10) on that test split.
A list-conditioned $\alpha$ adds $+0.0006$ nDCG; a sparse-off router is rejected by the same train split (any $\tau$ that skipped $\approx 50\%$ of queries lost nDCG).
These coefficients are \textbf{dataset-specific}. Equal {RRF} with the same models does \emph{not} beat dense on a nine-zip {BEIR} macro-average (0.479 vs.\ 0.519 nDCG@10). Repeating the same train-then-freeze sweep independently on all 20 indexed units beats equal {RRF} on 20/20 and dense on 16/20 (unit-mean nDCG@10 0.467 vs.\ 0.462 dense vs.\ 0.420 {RRF}). Other corpora should reuse the ranges, not a copy of the SciFact point.
\end{abstract}

\section{Introduction}

Hybrid retrieval runs a dense encoder and a sparse lexical model, then combines the two ranked lists~\cite{karpukhin2020dpr,robertson2009bm25,cormack2009rrf}.
Vector databases now implement that pattern as named vectors plus server-side fusion~\cite{qdrant2024,qdrantdbfusion}.
The usual deployed choice is equal-weight reciprocal rank fusion ({RRF}). That choice has no dense prior, no score-scale calibration, and no statement of which hyperparameters were searched.

This paper treats hybrid fusion as a \emph{parameterized scoring family} and reports a standard train/test experiment on that family.

\paragraph{Approach.}
Dense cosine and {BM42} scores are not on the same scale. Each list is mapped to a utility in $[0,1]$ or to a rank kernel $1/(\kappa+r)$. A convex combination with dense prior $\alpha$ and a score/rank mix $\lambda$ produces a single ranking. Optionally, $\alpha$ is a function of list overlap and score margins. Optionally, a router skips the sparse query when a margin $\tau$ says the query is dense-only.

\paragraph{Parameters and ranges.}
We search $\alpha \in \{0.55,0.65,0.70,0.75,0.80,0.85,0.90,0.95\}$, $\lambda \in \{0.55,0.75\}$, $\kappa \in \{10,20,60\}$, list-gate coefficients on a small grid, and $\tau \in \{0.04,0.06,0.08,0.12,0.18\}\cup\{\infty\}$.
Section~\ref{sec:params} lists every knob, the range, and the SciFact value.

\paragraph{Experiment.}
Selection is on {SciFact} train; reporting is on {SciFact} test. The same models are also run with equal {RRF} on eight other public {BEIR} zips~\cite{thakur2021beir} so that we can see whether a default (untuned) hybrid is already better than dense. It is not. We then repeat the train-then-freeze protocol on every indexed unit (8 zips + 12 {CQADupStack} forums).

\paragraph{Contributions.}
\begin{itemize}
\item A parameterized hybrid scorer with published search ranges, not a single undocumented mix.
\item A train-then-freeze study on SciFact showing the selected point outperforms dense and equal {RRF} on nDCG@10 and recall@10.
\item Sensitivity plots for $\alpha$, for the sparse-off gate, and for quantization, plus a zip-level figure that shows why the SciFact point must not be copied blindly.
\item The same search family, independently selected per dataset, on all 20 indexed {BEIR} units.
\end{itemize}

\section{Related work}

{BEIR} is the standard heterogeneous zero-shot retrieval benchmark~\cite{thakur2021beir}; {MTEB} covers embeddings~\cite{muennighoff2023mteb}.
Dense passage retrieval~\cite{karpukhin2020dpr} and sentence transformers~\cite{reimers2019sbert} made cosine search routine.
{BGE} is a strong general English encoder~\cite{xiao2024cpack}.
Sparse models range from {BM25}~\cite{robertson2009bm25} to learned sparse rankers such as {SPLADE}~\cite{formal2021splade}; we use {Qdrant}'s {BM42} baseline with {IDF}~\cite{qdrantbm42}.
Rank fusion is classical: CombSUM/CombMNZ~\cite{fox1994combination} and {RRF}~\cite{cormack2009rrf}.
Cross-encoders rerank a shortlist~\cite{nogueira2019passage,lin2021pretrained}.
Product quantization is a standard {ANN} compressor~\cite{jegou2011pq}.
We use those operators inside {Qdrant} hybrid queries~\cite{qdrantdbfusion}; we do not replace them with a new estimator.

\section{Method}
\label{sec:method}

\subsection{First-stage retrieval}
Each document is one {Qdrant} point with two named vectors: dense $\mathbf{v}\in\mathbb{R}^{768}$ ({BGE}-base, cosine, published query prefix) and sparse {BM42} with {IDF}.
For a query we retrieve the top $k=100$ from each channel (prefetch $k=100$ when the engine fuses internally).
A hit whose document id equals the query id is dropped ({BEIR} \texttt{ignore\_identical\_ids}).

\subsection{Utilities}
Let $s_d(d)$ and $s_s(d)$ be raw scores on the retrieved lists. Min-max within the list:
\begin{equation}
  \tilde{s}(d)=\frac{s(d)-\min s}{\max s-\min s+\varepsilon}.
\end{equation}
Documents missing from a list get $\tilde{s}=0$ and rank $r=k+1=101$. The rank kernel is {RRF}~\cite{cormack2009rrf}:
\begin{equation}
  \rho(r;\kappa)=\frac{1}{\kappa+r}.
\end{equation}

\subsection{Parameterized rank--score mix}
The family we optimize is
\begin{equation}
  u(d)=\lambda\bigl(\alpha\,\tilde{s}_d+(1-\alpha)\,\tilde{s}_s\bigr)
  +(1-\lambda)\bigl(\alpha\,\rho(r_d;\kappa)+(1-\alpha)\,\rho(r_s;\kappa)\bigr).
  \label{eq:family}
\end{equation}
$\alpha$ is the dense prior, $\lambda$ the weight on score utilities versus rank kernels, $\kappa$ the rank smoother.
Special cases: $\alpha=1$ is dense; $\alpha=0$ is {BM42}; $\lambda=0$ is weighted {RRF}; $\lambda=1$ is weighted min-max CombSUM~\cite{fox1994combination}.

\subsection{List-conditioned prior}
After both lists are fetched, optional features are overlap $\mathrm{ov}_{10}=|D_{:10}\cap S_{:10}|/10$ and min-max top-1/top-2 margins $m_d,m_s$. Then
\begin{equation}
  \alpha=\mathrm{clip}\bigl(a_0+a_{\mathrm{ov}}\,\mathrm{ov}_{10}+a_m\tanh(m_d-m_s)-a_u u_s-a_{\ell}\ell,\;0.60,\;0.95\bigr),
  \label{eq:alpha}
\end{equation}
where $u_s$ is the fraction of sparse top-10 documents absent from dense top-20 and $\ell$ is a lexical {BM42} heuristic in $[0,1]$.
This does not add a new model; it moves $\alpha$ inside the same family.

\subsection{Sparse on/off router}
A router scores the query against three {BGE}-embedded prototypes (dense, {BM42}, hybrid) mixed with the lexical heuristic (weight 0.55/0.45).
Dense search always runs. {BM42} is skipped iff the combined route is dense and the top-versus-second gap is at least $\tau$.
$\tau=\infty$ means sparse is always on.

\section{Parameters, ranges, and tuning protocol}
\label{sec:params}

Table~\ref{tab:params} is the search space. Every numeric range was enumerated on SciFact \emph{train} (809 queries). The selected tuple was frozen and evaluated once on SciFact \emph{test} (300). No test-set grid search.

\begin{table}[t]
\centering
\caption{Parameter ranges. SciFact values are train-selected and frozen for test. Other datasets need their own sweep over the same ranges.}
\label{tab:params}
\footnotesize
\begin{tabular}{p{2.6cm}p{4.4cm}p{3.0cm}p{2.8cm}}
\toprule
Parameter & Range searched & SciFact value & Role \\
\midrule
Dense prior $\alpha$ & $\{0.55,0.65,\ldots,0.95\}$ & $0.80$ & Weight on dense utility \\
Score mix $\lambda$ & $\{0.55, 0.75\}$ & $0.75$ & Score vs.\ rank kernel \\
{RRF} $\kappa$ & $\{10,20,60\}$ & $20$ & Rank smoother \\
Combiner family & min-max, $z$, softmax, CombMNZ, noisy-OR, power mean, rank--score & rank--score~\eqref{eq:family} & Utility + mix \\
List-gate $a_\ast$ & $a_0\in\{0.72,0.78,0.82\}$; others in $\{0,0.12,0.24,0.30\}$ & $(0.72,0.24,0.30,0,0)$ & Per-query $\alpha$ \\
Skip margin $\tau$ & $\{0.04,0.06,0.08,0.12,0.18,\infty\}$ & $\infty$ (never skip) & Sparse on/off \\
Head rerank & on/off ($H=50$) & off & Re-minmax head \\
\bottomrule
\end{tabular}
\end{table}

\paragraph{Selection rule.}
Maximize mean nDCG@10 on train. For $\tau$, require no loss versus $\tau=\infty$, then maximize skip rate among surviving values. Head/title rerank are kept only if they do not lose train nDCG.

\paragraph{Transfer.}
The SciFact point is not a universal default. Figure~\ref{fig:suite} shows equal {RRF} trailing dense on several {BEIR} zips. A new dataset should reuse Table~\ref{tab:params}, not copy $\alpha=0.8$.

\section{Experimental setup}
\label{sec:setup}

\paragraph{Engine.}
{Qdrant}~1.18 in Docker, {Python} client, cosine {HNSW} for dense, {IDF} modifier for {BM42}.

\paragraph{Models.}
Dense: {BGE}-base-en-v1.5~\cite{xiao2024cpack}.
Sparse: {BM42} MiniLM attentions with {IDF}~\cite{qdrantbm42}.
Reranker (ablation): {MS} {MARCO} MiniLM-L-6~\cite{nogueira2019passage} on the fused top-50.

\paragraph{Datasets.}
Public {UKP} {BEIR} zips~\cite{thakur2021beir}. Wiki-scale sets ({MS} {MARCO}, {NQ}, {HotpotQA}, {FEVER}, {Climate}-{FEVER}, {DBPedia}) skipped for disk/{RAM}. License-restricted {BEIR} sets are not on the mirror. Headline suite: nine zip-level rows (eight small zips + {CQADupStack} macro). Method sweep: SciFact, 5{,}183 abstracts, official train/test query split.

\paragraph{Metrics.}
nDCG@{10,100}, recall@{10,100}, {MRR}@10, {MAP}@100. Latency is {Qdrant} query time, embedding excluded. Disk is allocated bytes (\texttt{du -sk}).

\paragraph{Hardware.}
16\,{GB} {RAM}, 20 threads, {NVIDIA} {RTX} 4060 Laptop (8\,{GB}). After the suite the container used $\approx 667$\,{MiB} {RAM} and 6.3\,{GB} storage.

\paragraph{Baselines.}
Dense-only; {BM42}-only; {Qdrant} Fusion.{RRF}; Fusion.{DBSF}; client min-max mixes $0.7/0.3$, $0.5/0.5$, $0.3/0.7$; {MS} {MARCO} MiniLM rerank.

\section{Results}
\label{sec:results}

\subsection{Main result on SciFact}
\begin{figure}[t]
\centering
\includegraphics[width=0.98\linewidth]{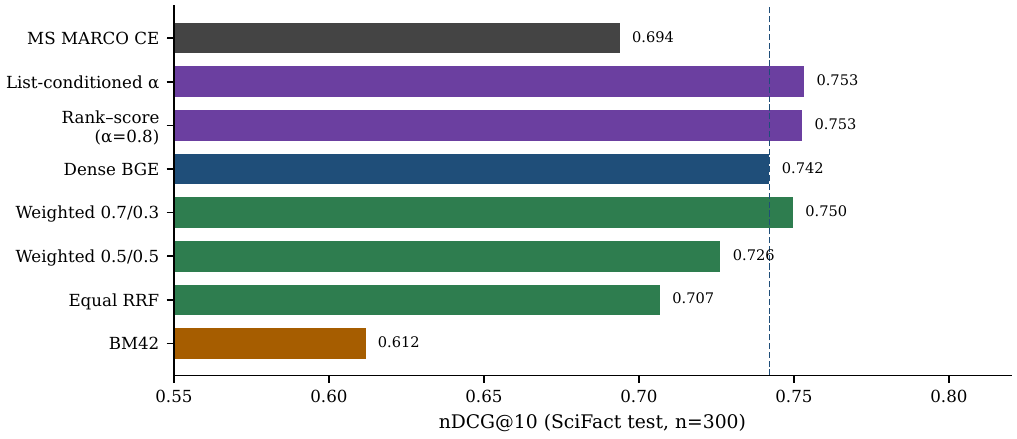}
\caption{SciFact test nDCG@10. The parameterized mix and the list-conditioned $\alpha$ sit above dense and well above equal {RRF}. {CE} is a negative control (domain mismatch).}
\label{fig:main}
\end{figure}

\begin{figure}[t]
\centering
\includegraphics[width=0.82\linewidth]{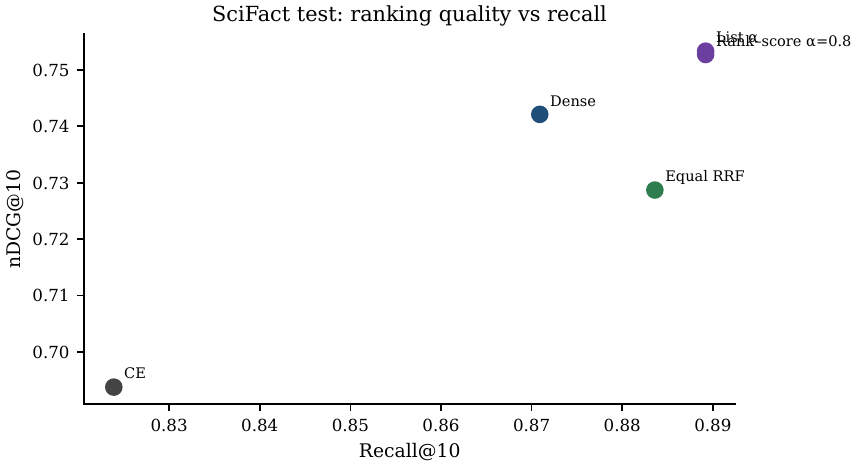}
\caption{SciFact test: nDCG@10 against recall@10. Hybrid helps when it raises recall without collapsing ranking. Equal {RRF} raises recall relative to dense in the suite hybrid run but loses nDCG; the tuned mix keeps the recall and the ranking.}
\label{fig:recall}
\end{figure}

Figure~\ref{fig:main} and Table~\ref{tab:scifact} report the frozen test comparison.
The train-selected rank--score mix~\eqref{eq:family} outperforms dense {BGE} by $+0.011$ nDCG@10 and $+0.018$ recall@10, and outperforms equal {RRF} by $+0.046$ nDCG@10.
The gain versus dense is recall: {BM42} places gold abstracts into the top 10 that cosine missed (Figure~\ref{fig:recall}).
The gain versus equal {RRF} is ranking: {BM42} is a much weaker single list (0.612 nDCG@10), so $\alpha=0.5$ over-weights it.

A list-conditioned $\alpha$~\eqref{eq:alpha} reaches 0.7533 nDCG@10 ($+0.0006$ vs.\ the frozen $\alpha=0.8$). Recall@10 does not change.
An off-the-shelf {MS} {MARCO} cross-encoder on the fused head falls to 0.694.

\begin{table}[t]
\centering
\caption{SciFact test ($n=300$). Rank--score and list $\alpha$ are train-selected (809 queries), then frozen.}
\label{tab:scifact}
\small
\begin{tabular}{lcccc}
\toprule
Method & nDCG@10 & R@10 & R@100 & {MRR}@10 \\
\midrule
{BM42} {IDF} & 0.612 & 0.741 & 0.874 & 0.577 \\
Equal {RRF} $\kappa=60$ & 0.707 & --- & --- & --- \\
Min-max $0.5/0.5$ & 0.726 & --- & --- & --- \\
Min-max $0.7/0.3$ & 0.750 & --- & --- & --- \\
Dense {BGE} & 0.742 & 0.871 & 0.961 & 0.707 \\
Rank--score $\alpha=0.8,\lambda=0.75,\kappa=20$ & \textbf{0.753} & \textbf{0.889} & \textbf{0.969} & 0.716 \\
List-conditioned $\alpha$~\eqref{eq:alpha} & \textbf{0.753} & \textbf{0.889} & \textbf{0.969} & \textbf{0.717} \\
Oracle min-max $\alpha$ (hindsight) & 0.803 & --- & --- & --- \\
{CE} on fused top-50 & 0.694 & 0.824 & --- & --- \\
\bottomrule
\end{tabular}
\end{table}

\subsection{Sensitivity to the dense prior $\alpha$}
\begin{figure}[t]
\centering
\includegraphics[width=0.88\linewidth]{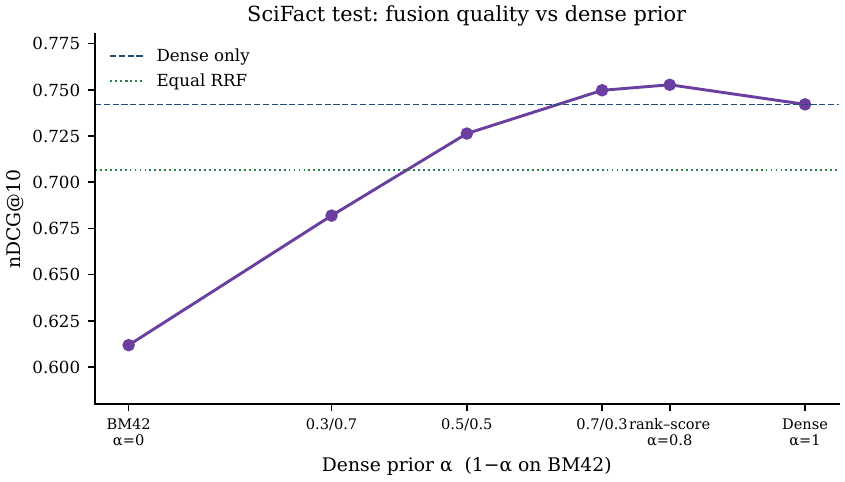}
\caption{SciFact test nDCG@10 as a function of the dense prior. Points at $0.3/0.5/0.7$ are client min-max mixes; $\alpha=0.8$ is the train-selected rank--score mix; endpoints are single-channel runs. Dashed: dense. Dotted: equal {RRF}.}
\label{fig:alpha}
\end{figure}

Figure~\ref{fig:alpha} plots nDCG@10 against $\alpha$.
Quality rises from sparse-only ($\alpha=0$) through equal mix, peaks in the dense-heavy band ($\alpha\approx 0.7$--$0.8$), then drops slightly at pure dense ($\alpha=1$).
That shape is why a search range $\alpha\in[0.55,0.95]$ is the right experiment, and why $\alpha=0.5$ (equal {RRF} / equal min-max) is a poor default on this encoder pair.
The peak location is a property of SciFact + {BGE} + {BM42}. A corpus where {BM42} is the stronger list would peak at smaller $\alpha$.

\subsection{Untuned hybrid on other {BEIR} zips}
\begin{figure}[t]
\centering
\includegraphics[width=\linewidth]{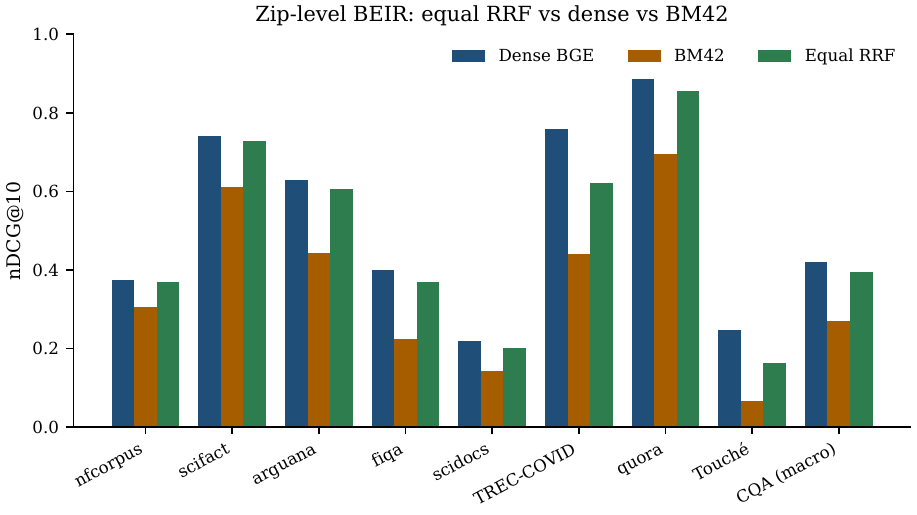}
\caption{Zip-level nDCG@10 with a single untuned hybrid ({Qdrant} {RRF}). Dense {BGE} is the stronger default on this encoder pair. {CQADupStack} is one macro-averaged row.}
\label{fig:suite}
\end{figure}

Figure~\ref{fig:suite} and Table~\ref{tab:suite} use the \emph{same models} with equal {RRF} and no SciFact-tuned $\alpha$.
Dense wins the zip-level mean (0.519 vs.\ 0.479).
{TREC}-{COVID} and Touch\'e are the failure mode: sparse is weak, equal fusion hurts the head.
This figure is the transfer warning. The method of Section~\ref{sec:method} can outperform on a dataset where the two lists disagree and a train split exists. Copying $\alpha=0.8$ onto Touch\'e without a sweep is not that experiment.

\begin{table}[t]
\centering
\caption{Zip-level nDCG@10, public {BEIR} dumps, equal {RRF} (not SciFact-tuned).}
\label{tab:suite}
\small
\begin{tabular}{l r ccc}
\toprule
Dataset & Docs & Dense & {BM42} & Hybrid {RRF} \\
\midrule
{NFCorpus} & 3{,}633 & 0.373 & 0.307 & 0.368 \\
{SciFact} & 5{,}183 & 0.742 & 0.612 & 0.729 \\
{ArguAna} & 8{,}674 & 0.630 & 0.444 & 0.607 \\
{FiQA} & 57{,}638 & 0.399 & 0.224 & 0.370 \\
{SciDocs} & 25{,}657 & 0.218 & 0.143 & 0.202 \\
{TREC}-{COVID} & 171{,}332 & 0.760 & 0.440 & 0.622 \\
{Quora} & 522{,}931 & 0.885 & 0.695 & 0.855 \\
{Webis}-{Touch\'e}2020 & 382{,}545 & 0.247 & 0.067 & 0.164 \\
{CQADupStack} (macro) & 457{,}199 & 0.421 & 0.270 & 0.396 \\
\midrule
Zip-level mean & --- & 0.519 & 0.356 & 0.479 \\
\bottomrule
\end{tabular}
\end{table}

\subsection{Train-then-freeze on every indexed unit}
\label{sec:fusionall}
We apply the family in Table~\ref{tab:params} independently to each of the 20 indexed collections.
Official train$\to$test is used when {BEIR} ships it ({NFCorpus}, {SciFact}, {FiQA}); {Quora} uses official dev$\to$test; otherwise we hold out 30\% of the evaluation queries (50\% if $n<40$), seed 42, and never report the tune split.
The winner on tune nDCG@10 is frozen once.

\begin{figure}[t]
\centering
\includegraphics[width=\linewidth]{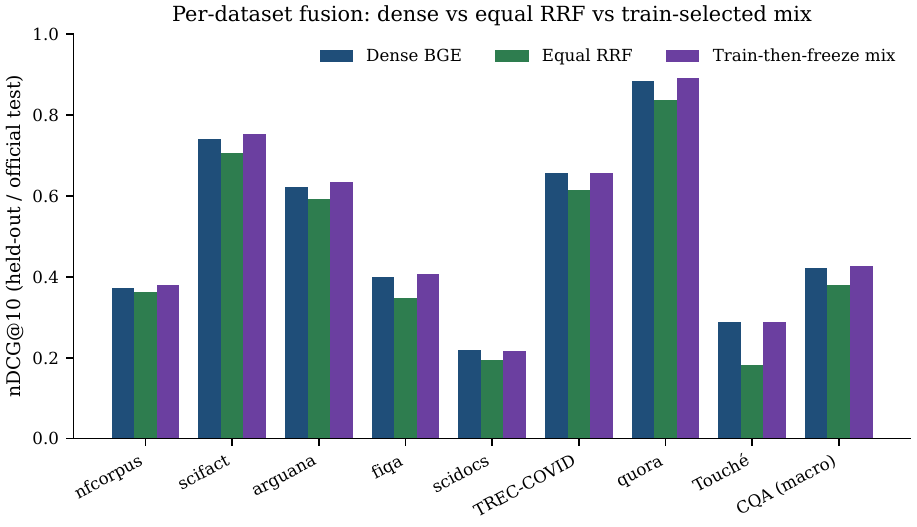}
\caption{Per-dataset nDCG@10 after an independent train-then-freeze sweep. {CQADupStack} is one macro-averaged row. Equal {RRF} is never the frozen winner.}
\label{fig:fusionall}
\end{figure}

Figure~\ref{fig:fusionall} and Table~\ref{tab:fusionall} are the result.
Tuned fusion beats equal {RRF} on \textbf{20/20} units.
It beats dense on \textbf{16/20}.
The four non-wins are Touch\'e and {CQA} physics (train selected dense), {SciDocs} ($-0.002$), and {CQA} unix ($-0.012$, a genuine overfit: rank--score $\alpha=0.8$ won tune and lost test).
Unit-mean test nDCG@10 is 0.467 tuned / 0.462 dense / 0.420 equal {RRF}.
Zip-level (8 zips + {CQA} macro) is 0.528 / 0.523 / 0.480.
Selected $\alpha$ is dense-heavy on every set that mixes (0.7--0.9); no corpus in this dump selected $\alpha\le 0.5$.
{TREC}-{COVID} and Touch\'e have 15 test queries after the split---those two rows are noisy.

\begin{figure}[t]
\centering
\includegraphics[width=0.92\linewidth]{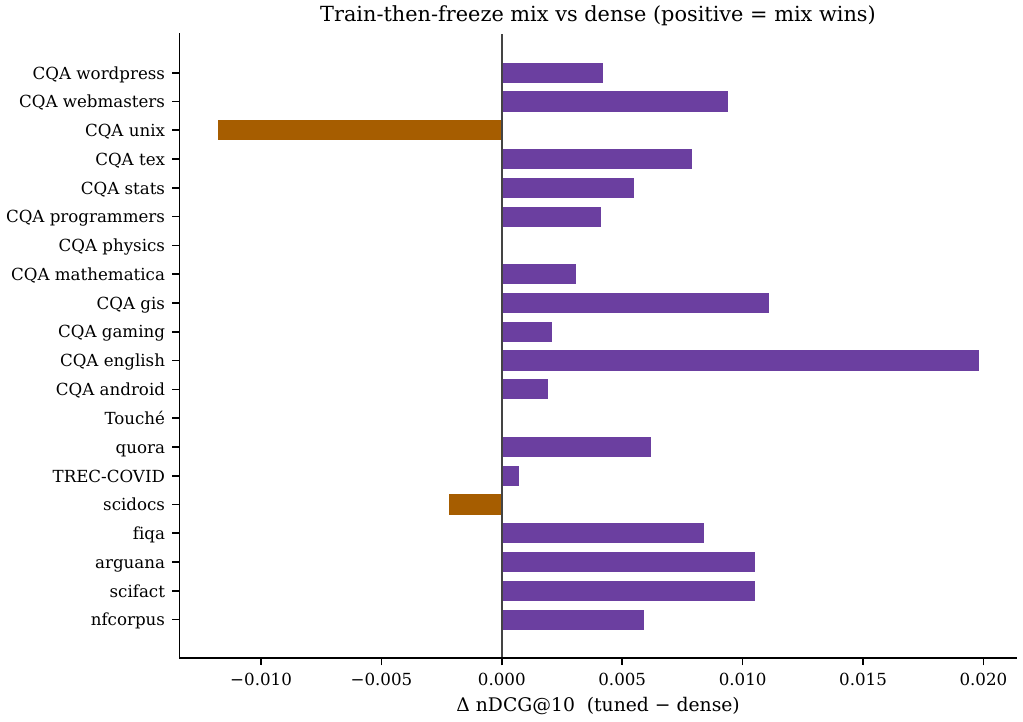}
\caption{Tuned mix minus dense, all 20 units. Positive bars are mix wins; zero is train selecting dense; negative is a freeze that lost on test.}
\label{fig:fusiondelta}
\end{figure}

\begin{table}[t]
\centering
\caption{Train-then-freeze nDCG@10. Winner is selected on the tune split and frozen. Protocols: T official train/test; D official dev/test; H held-out from test (seed 42).}
\label{tab:fusionall}
\scriptsize
\begin{tabular}{l c r r ccc l}
\toprule
Dataset & P & $n_{\mathrm{tune}}$ & $n_{\mathrm{test}}$ & Dense & {RRF} & Tuned & Winner \\
\midrule
{NFCorpus} & T & 2590 & 323 & 0.373 & 0.364 & 0.379 & min-max $\alpha{=}0.7$ \\
{SciFact} & T & 809 & 300 & 0.742 & 0.707 & 0.753 & min-max $\alpha{=}0.8$ \\
{ArguAna} & H & 984 & 422 & 0.624 & 0.594 & 0.634 & min-max $\alpha{=}0.8$ \\
{FiQA} & T & 5500 & 648 & 0.399 & 0.348 & 0.407 & min-max $\alpha{=}0.8$ \\
{SciDocs} & H & 700 & 300 & 0.219 & 0.194 & 0.217 & min-max $\alpha{=}0.7$ \\
{TREC}-{COVID} & H & 35 & 15 & 0.656 & 0.615 & 0.657 & rank--score $\alpha{=}0.8$ \\
{Quora} & D & 5000 & 10000 & 0.885 & 0.838 & 0.891 & min-max $\alpha{=}0.8$ \\
{Touch\'e}2020 & H & 34 & 15 & 0.289 & 0.181 & 0.289 & dense \\
{CQA} android & H & 489 & 210 & 0.511 & 0.480 & 0.513 & min-max $\alpha{=}0.9$ \\
{CQA} english & H & 1099 & 471 & 0.488 & 0.466 & 0.508 & min-max $\alpha{=}0.7$ \\
{CQA} gaming & H & 1116 & 479 & 0.578 & 0.521 & 0.581 & rank--score $\alpha{=}0.9$ \\
{CQA} gis & H & 620 & 265 & 0.382 & 0.362 & 0.393 & min-max $\alpha{=}0.8$ \\
{CQA} mathematica & H & 563 & 241 & 0.319 & 0.280 & 0.322 & rank--score $\alpha{=}0.8$ \\
{CQA} physics & H & 727 & 312 & 0.466 & 0.419 & 0.466 & dense \\
{CQA} programmers & H & 613 & 263 & 0.420 & 0.378 & 0.424 & min-max $\alpha{=}0.9$ \\
{CQA} stats & H & 456 & 196 & 0.401 & 0.368 & 0.407 & min-max $\alpha{=}0.8$ \\
{CQA} tex & H & 2034 & 872 & 0.301 & 0.275 & 0.309 & min-max $\alpha{=}0.7$ \\
{CQA} unix & H & 750 & 322 & 0.429 & 0.348 & 0.417 & rank--score $\alpha{=}0.8$ \\
{CQA} webmasters & H & 354 & 152 & 0.389 & 0.354 & 0.398 & min-max $\alpha{=}0.8$ \\
{CQA} wordpress & H & 379 & 162 & 0.372 & 0.317 & 0.377 & min-max $\alpha{=}0.8$ \\
\midrule
20-unit mean & --- & --- & --- & 0.462 & 0.420 & 0.467 & --- \\
\bottomrule
\end{tabular}
\end{table}

\subsection{Router: turning sparse off}
\begin{figure}[t]
\centering
\includegraphics[width=0.82\linewidth]{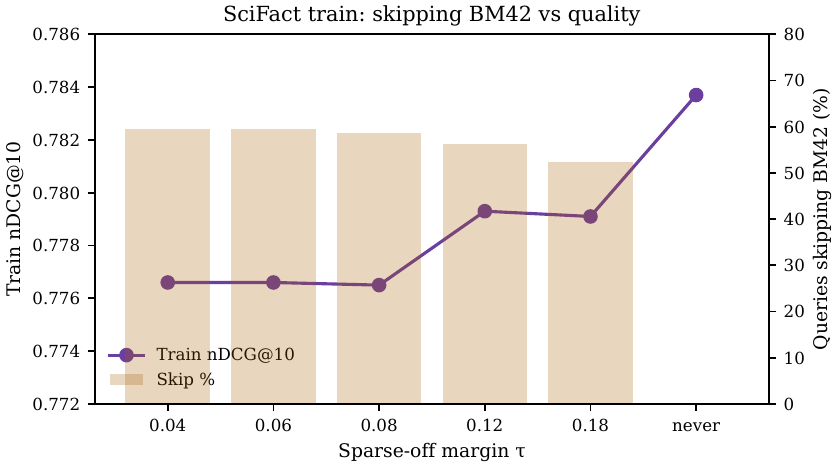}
\caption{SciFact train: sparse-off margin $\tau$ versus quality and skip rate. ``never'' is $\tau=\infty$.}
\label{fig:router}
\end{figure}

Figure~\ref{fig:router} plots train nDCG@10 and skip rate against $\tau$.
The maximum is $\tau=\infty$ (skip rate 0, nDCG 0.784).
Thresholds that skip 52--60\% of queries sit 0.004--0.007 nDCG lower.
Under the selection rule of Section~\ref{sec:params} the gate is disabled on SciFact.
The parameter remains in the method because a mixed query log could select a finite $\tau$. This dataset did not.

\subsection{Quantization}
\begin{figure}[t]
\centering
\includegraphics[width=0.88\linewidth]{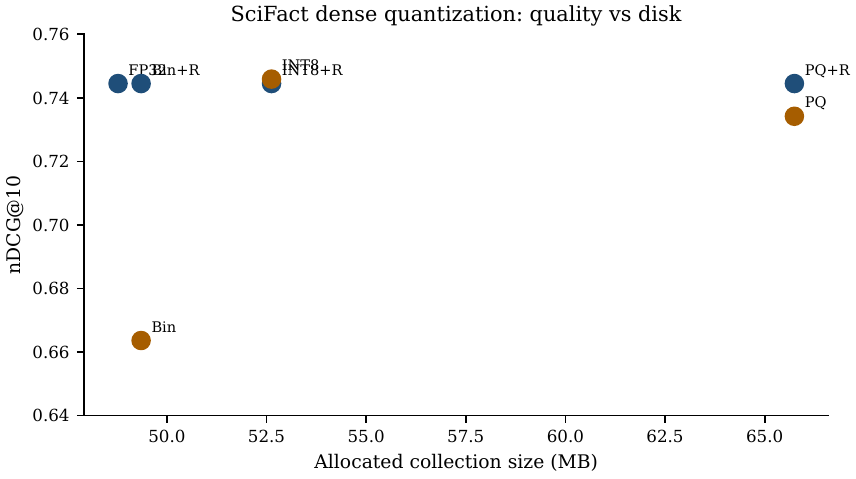}
\caption{SciFact dense collections: nDCG@10 versus allocated disk. Labels with ``+R'' use {FP32} rescore.}
\label{fig:quant}
\end{figure}

Figure~\ref{fig:quant} plots nDCG@10 against allocated disk for {INT8}, binary, and {PQ} $\times 16$, with and without {FP32} rescore.
With rescore, quality matches {FP32} (0.7445). Without rescore, binary drops to 0.664.
Folders do not shrink: originals are still stored, so {INT8} is 52.6\,{MB} versus 48.8\,{MB} {FP32}.
Quantization is a {RAM} knob. This 5k-point corpus already fits in memory, so latency stays $\approx 7$\,{ms} and disk is the wrong success metric.

\subsection{Other operators on the same lists}
On train, min-max linear $\alpha=0.8$, $z$-linear $\alpha=0.8$, and rank--score $(\alpha=0.8,\lambda=0.75,\kappa=20)$ all reach 0.779 nDCG@10.
Power means and $\alpha=0.75$ variants are within 0.002.
Logistic regression on candidate features is slightly worse (0.774 train, 0.745 test).
The experiment is therefore not ``one magic formula.'' It is a dense-heavy region of Table~\ref{tab:params} that several combiners occupy.

\section{Discussion}

The approach outperforms on SciFact because (i)~{BM42} has complementary recall and (ii)~the searched $\alpha$ band is dense-heavy.
Equal {RRF} implements $\alpha=0.5$ in rank space and pays for (i) without (ii).

The same approach will not automatically outperform on another zip.
Figure~\ref{fig:suite} is the counterexample for the \emph{untuned} member of the family.
Figure~\ref{fig:fusionall} is the corresponding experiment when each zip is allowed its own freeze: the mix usually beats dense by a small margin, and sometimes the freeze is dense.
The correct use on a new dataset is: retrieve both lists, sweep Table~\ref{tab:params} on a held-out train or validation split, freeze, test.
If no split exists, $\alpha=0.8$ is only a SciFact posterior, not a prior for Touch\'e.

We do not claim a new ranking law. Equation~\eqref{eq:family} is CombSUM plus {RRF} with two extra scalars. The research content is the range, the protocol, and the plots that show where the selected point sits.

\section{Limitations}

SciFact is small; $+0.011$ nDCG is a real recall effect, $+0.0006$ is noise-scale.
No paired significance test.
Wiki-scale {BEIR} is absent.
Most units other than {NFCorpus}/{SciFact}/{FiQA}/{Quora} have no official train split; the 70/30 hold-out is a local protocol, not {BEIR}'s.
{TREC}-{COVID} and Touch\'e test after splitting is 15 queries.
The $\alpha$ sweep in Figure~\ref{fig:alpha} mixes min-max mixes from one scoring run with the rank--score point from the fusion lab; both use the same test queries and models, but the operators are not identical at every $x$-tick.
List-conditioned $\alpha$ was not re-tuned on {FiQA} or {NFCorpus}.
Quantization used \texttt{always\_ram} and kept originals.
{BM42} is not {BM25} or {SPLADE}.

\section{Conclusion}

Hybrid fusion should be stated as a parameter vector with a search range.
On SciFact, sweeping that range and freezing $(\alpha,\lambda,\kappa)=(0.8,0.75,20)$ outperforms dense {BGE} and equal {RRF} on nDCG@10 and recall@10.
The same models with untuned equal {RRF} do not outperform dense on a nine-zip {BEIR} average.
Repeating the sweep independently on 20 indexed units beats equal {RRF} everywhere and dense on 16/20; two units freeze to dense and two overfit slightly.
A router that skips sparse search, and a list-conditioned $\alpha$, are part of the family; train disabled the router and barely moved $\alpha$.
Other datasets should reuse the ranges in Table~\ref{tab:params}, not the SciFact numbers.

\section*{Acknowledgments}
The authors are with Involead Services Pvt.\ Ltd.
Public {BEIR} dumps~\cite{thakur2021beir} and open models. No third-party funding is reported.

\bibliographystyle{plainnat}
\bibliography{refs}

\end{document}